\documentclass[journal,transmag]{IEEEtran}
\ifCLASSINFOpdf
\else
\fi
\usepackage[numbers, sort&compress]{natbib}
\usepackage{amsmath,amssymb,amsfonts}
\usepackage{dblfloatfix} 
\usepackage{graphicx}
\usepackage{xcolor}

\begin{document}
%
\title{Effect of Single-Ion Anisotropy on Stability of Quantum and Thermal Entanglement in a Mixed-Spin Heisenberg Trimer}


\author{\IEEEauthorblockN{H. Vargov\'a\IEEEauthorrefmark{1} and
J. Stre\v{c}ka\IEEEauthorrefmark{2}}
\IEEEauthorblockA{\IEEEauthorrefmark{1}Institute of Experimental Physics, Slovak Academy of Sciences, Watsonova 47, 040 01 Ko\v {s}ice, Slovakia}
\IEEEauthorblockA{\IEEEauthorrefmark{2}Department of Theoretical Physics and Astrophysics, Faculty of Science, Pavol  Jozef \v{S}af\'{a}rik University,\\ Park Angelinum 9, 040 01 Ko\v{s}ice, Slovakia}
\thanks{Corresponding author: H. Vargov\'a (email: hcencar@saske.sk).}}

\markboth{IEEE Transactions on Magnetics, The Program Code: 6P-07, CSMAG'25}%
{Shell \MakeLowercase{\textit{et al.}}: Bare Demo of IEEEtran.cls for IEEE Transactions on Magnetics Journals}

%



\IEEEtitleabstractindextext{%
\begin{abstract}
The effect of uniaxial single-ion anisotropy on quantum entanglement is rigorously quantified using negativity in a mixed spin-($1$,$1/2$,$1$) Heisenberg trimer, accounting for different exchange coupling constants between identical and distinct spins. Bipartite negativities between the single-spin entity and the remaining spin dimer are analyzed alongside the global tripartite negativity (gTN) of the whole trimer under the effect of an external magnetic field and both easy-axis and easy-plane types of single-ion anisotropy. Interestingly, the single-ion anisotropy significantly influences the degree of entanglement by altering the stability regions of energetically preferred ground states and it may also introduce additional phases in the overall ground-state phase diagram. Moreover, it is demonstrated that within specific ground states, the degree of entanglement primarily depends on the strength of single-ion anisotropy, altering the respective probability amplitudes of the corresponding eigenvectors. Finally, the thermal stability of entanglement is discussed in detail, including the emergence of a peculiar local minimum at finite temperatures. The obtained theoretical results may offer deeper insights into bipartite and tripartite  entanglement in trimetallic Ni$^{2+}$-Cu$^{2+}$-Ni$^{2+}$ molecular compounds.
\end{abstract}

}

\maketitle

\IEEEdisplaynontitleabstractindextext

%
\IEEEpeerreviewmaketitle

\section{Introduction}

\IEEEPARstart{Q}{uantum} entanglement~\cite{Horodecki} is a fundamental feature of composite quantum systems, wherein knowledge of one particle's state immediately determines the state of its entangled partner(s) without the need for additional measurement.
Owing to this property, entanglement has become a cornerstone of quantum science and novel technologies~\cite{Nielsen,Furusawa}. While a bipartite entanglement is well understood~\cite{Horodecki}, gaining deeper insight into a global entanglement in many-body systems remains a significant challenge~\cite{Shi}. A suitable theoretical platform for studying minimal yet nontrivial multipartite correlations is the Heisenberg trimer, where inherent asymmetries can strongly affect the degree of global tripartite entanglement (gTE) (see Refs.~\cite{Zad2016,Ghannadan2025,Vargova2025} and references therein).
In this study, we aim to advance the understanding of this issue by examining the stability of gTE in a mixed spin-($1$,$1/2$,$1$) Heisenberg trimer, which can be experimentally realized  in molecular nanomagnets with a Ni$_2$Cu magnetic core~\cite{Ribas,Podlesnyak}. Our analysis specifically focuses on the role of uniaxial single-ion anisotropy, commonly present in real magnetic compounds with higher spins. 
 
\section{Model and Methods}
Let us consider the Hamiltonian  
\allowdisplaybreaks
\begin{align}
\hat{\cal H}&\!=\!\!\sum_{i=1,2}\!\!\left(J\hat{\boldsymbol\mu}\cdot\hat{\bf S}_{i}\!+\!D(\hat{S}^{z}_{i})^2
\!-\!h\hat{S}^{z}_{i}\right)
\!+\! J_1\hat{\bf S}_{1}\cdot\hat{\bf S}_{2}
\!-\!h\hat{\mu}^{z}.
\label{eq1}
\end{align} 
where Heisenberg spins, $\mu=1/2$ and $S_1=S_2=1$,  interact with each other via exchange coupling constants $J$ and $J_1$. Here, $h=g\mu_BB^z$ represents the effect of an external magnetic field $B^z$ applied along the $z$-direction,  $g$ is the Land\'e $g$-factor and $\mu_B$ is the Bohr magneton. The parameter $D$ denotes the uniaxial single-ion anisotropy acting only on the spin-$1$ ions,  which can be either positive (the easy-plane type) or negative (the easy-axis type). For simplicity, we set the interaction $J$ as the energy unit.

The quantum bipartite entanglement of a tripartite system $ABC$ is quantified using bipartite negativity, defined as ${\cal N}_{A|BC}=\sum_{\lambda_i<0} |\lambda_i|$, where the sum runs over the absolute values of all negative eigenvalues $\lambda_i$ of the  density matrix $\rho_{ABC}^{T_A}$ that is partially transposed with respect to the subsystem $A$~\cite{Vidal}. The total density matrix is calculated using the standard expression, $\hat{\rho}\!=\!\tfrac{1}{\cal Z}\sum_i {\rm e}^{-\beta \epsilon_i}\vert \psi_i\rangle \langle \psi_i\vert$, where ${\cal Z}$ is the partition function and $\varepsilon_i$ ($\vert \psi_i\rangle$) are the eigenvalues (eigenvectors) of the Hamiltonian~\eqref{eq1}.
The subscript $A|BC$ refers to the bipartite entanglement between the spin $A$ and the spin dimer $BC$, which is evaluated by performing a partial transposition with respect to the subsystem $A$. There exist two other permutations with different individual spin. Due to the model's symmetry, only two distinct bipartite negativities are relevant: ${\cal N}_{\mu|S_1S_2}$ and ${\cal N}_{S_1|\mu S_2}$. The gTN is then calculated as the geometric mean of all three bipartite contributions, ${\cal N}_{ABC}=\sqrt[3]{{\cal N}_{A|BC}{\cal N}_{B|AC}{\cal N}_{C|AB}}$~\cite{Sabin}. The full details of all calculations are presented in Ref.~\cite{Vargova2024}.

The eigenvalues $\varepsilon_{S_t,S_t^z}$ and eigenvectors $\vert S_t,S_t^z\rangle$ of the model~\eqref{eq1} are categorized according to the eigenvalues of the composite operators: $\hat{S}_t\!=\!\hat{\mu}\!+\!\hat{S}_1\!+\!\hat{S}_2$ and its $z$-component $\hat{S}^z_t$. The complete list is collected in Tab.~\ref{tab1}.
\allowdisplaybreaks
\begin{table*}[h!]
 \caption{The list of eigenvalues $\varepsilon_{S_t,S_t^z}$ and eigenvectors $\vert S_t,S_t^z\rangle$ of the model~\eqref{eq1}. Here  $\vert \Phi_A^{\pm}\rangle\!=\!\tfrac{1}{\sqrt{2}}[\vert\!\pm1,0\rangle\!-\!\vert0,\!\pm1\rangle]$, $\vert \Phi_S^{\pm}\rangle\!=\!\tfrac{1}{\sqrt{2}}[\vert\!\pm1,0\rangle\!+\!\vert0,\!\pm1\rangle]$, and $\vert \Psi^{\pm}\rangle\!=\!\tfrac{1}{\sqrt{2}}[\vert\!\pm1,\!\mp1\rangle\!-\!\vert\!\mp1,\pm1\rangle]$,  
}
\resizebox{1.0\textwidth}{!}{  
\begin{tabular}{l|l|c}
 Eigenvalues $\varepsilon_{S_t,S_t^z}$ &  Eigenvectors $\vert S_t,S_t^z\rangle$ & Used abbreviation\\
\hline
$\varepsilon_{\tfrac{5}{2},\pm\tfrac{5}{2}}\!=\!\tfrac{1}{4}(6D\!-\!J)\!\mp\!\tfrac{5}{2}h\!+\!J_1\!+\!\frac{1}{4}(5J\!+\!2D)$, &$\vert \tfrac{5}{2},\pm\tfrac{5}{2}\rangle=\vert \!\pm\!\tfrac{1}{2}\rangle\otimes\vert\!\pm\!1,\!\pm\!1\rangle$,&$P_1^2\!=\!(5J\!+\!2D)^2\!-\!32JD$, $P_2^2\!=\!(3J\!-\!2D)^2\!+\!16JD$
 \\
$\varepsilon_{\tfrac{3}{2},\pm\tfrac{3}{2}}^{\rm I}\!=\!\tfrac{1}{4}(6D\!-\!J)\!\mp\!\tfrac{3}{2}h\!-\!J_1\!+\!\frac{1}{4}(3J\!-\!2D)$, & $\vert \tfrac{3}{2},\pm\tfrac{3}{2}\rangle^{\rm I}=\vert \!\pm\!\tfrac{1}{2}\rangle\otimes\vert\Phi^{\pm}_A\rangle$,&$Y_i\!=\!2{\rm sgn}(q)\sqrt{p}\cos(\tfrac{\phi}{3}\!+\!\tfrac{2}{3}\pi(3\!-\!i))$, ($i=0,1,2$)
\\
$\varepsilon_{\tfrac{3}{2},\pm\tfrac{3}{2}}^{\rm II}\!=\!\tfrac{1}{4}(6D\!-\!J)\!\mp\!\tfrac{3}{2}h\!+\!J_1\!-\!\frac{1}{4}P_1$,&$\vert \tfrac{3}{2},\pm\tfrac{3}{2}\rangle^{\rm II}=a_{1}^-\vert \!\pm\!\tfrac{1}{2}\rangle\otimes\vert\Phi^{\pm}_S\rangle\!+\!a_{1}^+\vert \!\mp\!\tfrac{1}{2}\rangle\otimes\vert \!\pm\!1,\!\pm\!1\rangle$,&$p\!=\!\tfrac{1}{36}(19J^2\!+\!12D^2)\!-\!\tfrac{J_1}{6}(J\!+\!2D\!-\!6J_1)$
\\
 $\varepsilon_{\tfrac{5}{2},\pm\tfrac{3}{2}}\!=\!\tfrac{1}{4}(6D\!-\!J)\!\mp\!\tfrac{3}{2}h\!+\!J_1\!+\!\frac{1}{4}P_1$,&$\vert \tfrac{5}{2},\pm\tfrac{3}{2}\rangle\;\;=\;a_{1}^+\vert \!\pm\!\tfrac{1}{2}\rangle\otimes\vert\Phi^{\pm}_S\rangle\!-\!a_{1}^-\vert \!\mp\!\tfrac{1}{2}\rangle\otimes\vert \!\pm\!1,\!\pm\!1\rangle$,& $q\!=\!(\tfrac{J}{3}-J_1)^3\!+\!(\tfrac{J}{3}-J_1)(D^2\!-\!DJ_1\!+\!\tfrac{3}{2}J_1J\!-\!J^2)\!+\!\tfrac{J^2}{4}(6J_1\!-\!D)$
\\
$\varepsilon_{\tfrac{1}{2},\pm\tfrac{1}{2}}^{\rm II}\!=\!\tfrac{1}{4}(6D\!-\!J)\!\mp\!\tfrac{1}{2}h\!-\!J_1\!-\!\frac{1}{4}P_2$,&$\vert \tfrac{1}{2},\pm\tfrac{1}{2}\rangle^{\rm II}=a^{-}_{2}\vert \!\pm\!\tfrac{1}{2}\rangle\otimes\vert\Psi^{\pm}\rangle\!+\!a^{+}_{2}\vert \!\mp\!\tfrac{1}{2}\rangle\otimes\vert\Phi^{\pm}_A\rangle$,  & $\phi\!=\!{\rm atan}(\sqrt{(p^3\!-\!q^2)/q})$
\\
 $\varepsilon_{\tfrac{3}{2},\pm\tfrac{1}{2}}^{\rm I}\!=\!\tfrac{1}{4}(6D\!-\!J)\!\mp\!\tfrac{1}{2}h\!-\!J_1\!+\!\frac{1}{4}P_2$,&
$\vert \tfrac{3}{2},\pm\tfrac{1}{2}\rangle^{\rm I}\;=a^{+}_{2}\vert \!\pm\!\tfrac{1}{2}\rangle\otimes\vert\Psi^{\pm}\rangle\!-\!a^{-}_{2}\vert \!\mp\!\tfrac{1}{2}\rangle\otimes\vert\Phi^{\pm}_A\rangle$,&$a_{1}^{\mp}=\mp\tfrac{\sqrt{P_1\mp(3J-2D)}}{\sqrt{2P_1}}$, $a^{\mp}_{2}=\tfrac{\sqrt{P_2\mp(J+2D)}}{\sqrt{2P_2}}$
\\
$\varepsilon_{\tfrac{5}{2},\pm\tfrac{1}{2}}\!=\!\tfrac{1}{6}(6D\!-\!J)\!\mp\!\tfrac{1}{2}h\!+\!Y_0$,&  $\vert \tfrac{5}{2},\pm\tfrac{1}{2}\rangle\;\;\;=\alpha_1\vert \!\pm\!\tfrac{1}{2}\rangle\otimes\vert\Psi^{\pm}\rangle\!+\!\vert \!\mp\!\tfrac{1}{2}\rangle\otimes[\beta_1\vert\Phi^{\pm}_A\rangle\!+\!\gamma_1\vert \!0,\!0\rangle]$,&$\alpha_{i}\!=\!(1\!+\!R_{i}^2\!+\!T_{i}^2/2)^{-1/2}$, $\beta_{i}\!=\!\alpha_{i}R_{i}$, $\gamma_{i}\!=\!\alpha_{i}T_{i}/\sqrt{2}$
\\
$\varepsilon_{\tfrac{3}{2},\pm\tfrac{1}{2}}^{\rm II}\!=\!\tfrac{1}{6}(6D\!-\!J)\!\mp\!\tfrac{1}{2}h\!+\!Y_1$,& $\vert \tfrac{3}{2},\pm\tfrac{1}{2}\rangle^{\rm II}=\alpha_2\vert \!\pm\!\tfrac{1}{2}\rangle\otimes\vert\Psi^{\pm}\rangle\!+\!\vert \!\mp\!\tfrac{1}{2}\rangle\otimes[\beta_2\vert\Phi^{\pm}_A\rangle\!+\!\gamma_2\vert \!0,\!0\rangle]$,  &$R_i\!=\!\tfrac{-\sqrt{2}}{J}(2Y_i\!-\!D\!-\!J_1\!-\!\tfrac{J}{6}\!+\!\tfrac{4DJ_1}{D\!-\!\tfrac{J}{6}\!+\!2J_1\!+\!2Y_i})$
\\
$\varepsilon_{\tfrac{1}{2},\pm\tfrac{1}{2}}^{\rm I}\!=\!\tfrac{1}{6}(6D\!-\!J)\!\mp\!\tfrac{1}{2}h\!+\!Y_2$,& $\vert \tfrac{1}{2},\pm\tfrac{1}{2}\rangle^{\rm I}\;=\alpha_3\vert \!\pm\!\tfrac{1}{2}\rangle\otimes\vert\Psi^{\pm}\rangle\!+\!\vert \!\mp\!\tfrac{1}{2}\rangle\otimes[\beta_3\vert\Phi^{\pm}_A\rangle\!+\!\gamma_3\vert \!0,\!0\rangle]$,  &$T_i\!=\!4(1\!+\!2\tfrac{(D\!-\!2J_1)}{(D\!-\!\tfrac{J}{6}\!+\!2J_1\!+\!2Y_i)})$
\end{tabular}
}
\label{tab1}
\end{table*}

\section{Results and discussion}
\begin{figure*}[b!]
\centering
{\includegraphics[width=5.8cm,height=4.53cm,trim=3.4cm 8.8cm 5.8cm 8.5cm, clip]{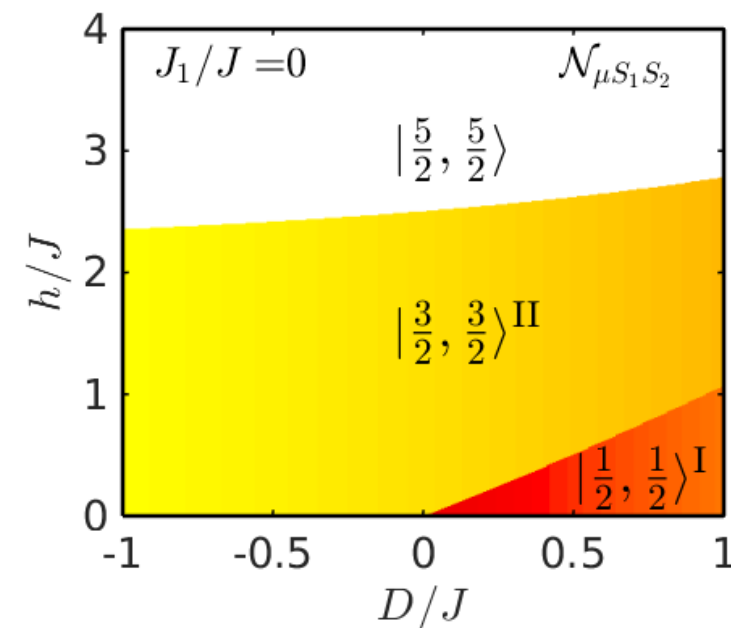}}
{\includegraphics[width=4.9cm,height=4.53cm,trim=5.35cm 8.8cm 5.8cm 8.5cm, clip]{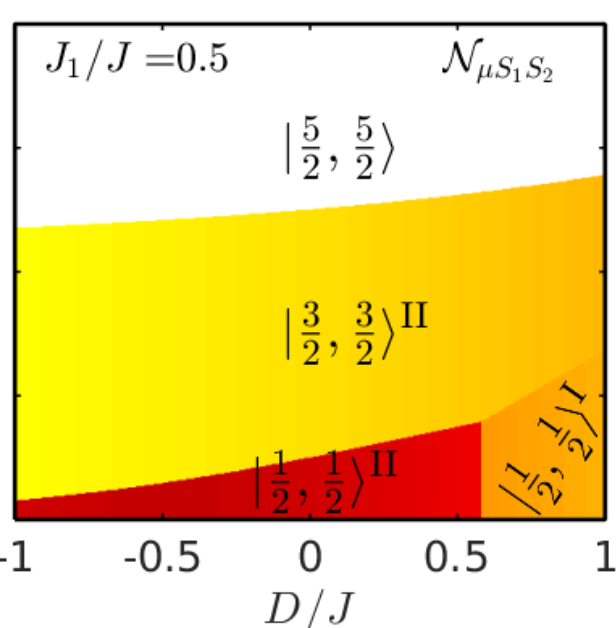}}
{\includegraphics[width=6.3cm,height=4.53cm,trim=5.36cm 8.8cm 3cm 8.5cm, clip]{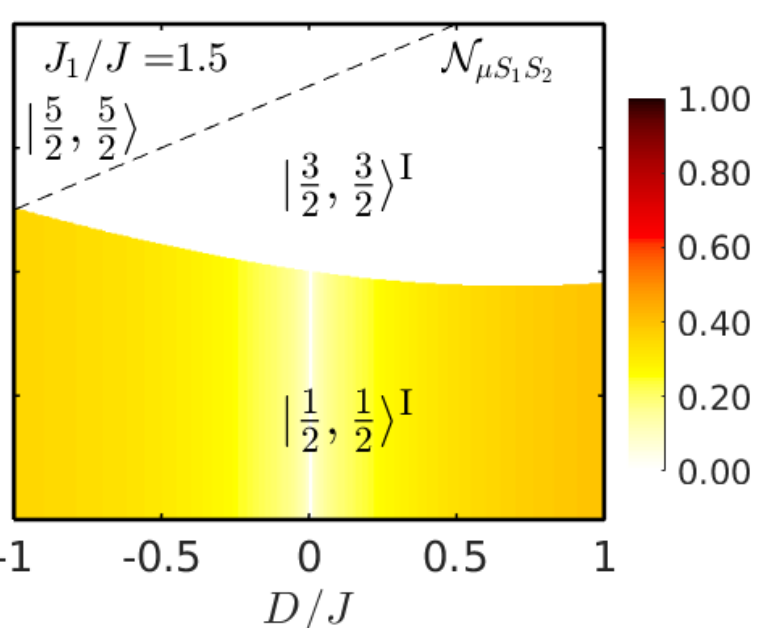}}
\caption{Density plots of the gTN as a function of $h/J$ and $D/J$ for three selected values of $J_1/J$.
}
\label{fig1}
\end{figure*}
To reasonably reduce the large parameter space, all subsequent discussions are restricted to specific ratios of the coupling constants $J_1/J\!=\!0$, $0.5$ and $1.5$. This choice is supported by our study of the same model in the limit of $D/J\!=\!0$, where three distinct  regimes were identified~\cite{Vargova2024}. Moreover, the case $J_1/J\!=\!0$ effectively models real complexes with nearly linear arrangements~\cite{Ribas,Podlesnyak}.

Fig.~\ref{fig1} illustrates the behavior of the gTN  as a function of $D/J$ and $h/J$ for selected values of $J_1/J$. The  anisotropy term $D/J$ significantly affects the strength of gTE, which may either increase or decrease depending on the favored ground state. This effect arises from the respective dependence of the probability amplitudes of the ground-state eigenvectors on the anisotropy term $D/J$. Moreover,  the uniaxial anisotropy $D/J\!\neq\!0$ notably shifts the transition magnetic field, i.e., the field at which the system undergoes a transition between ground states with different degrees of gTE.  Additionally, at smaller values of $J_1/J$, we identify the  $\vert \tfrac{1}{2},\tfrac{1}{2}\rangle^{\rm I}$ phase, previously observed only for $J_1/J\!\geq\!1$~\cite{Vargova2024}. This phase emerges from the competition between the easy-plane anisotropy $D/J\!>\!0$, which favors spin orientation in the $x$-$y$ plane, and the exchange couplings $J$ and $J_1$.

\begin{figure*}[t!]
{\includegraphics[width=0.33\textwidth,height=4.39cm,trim=3.2cm 8.5cm 4.3cm 8.9cm, clip]{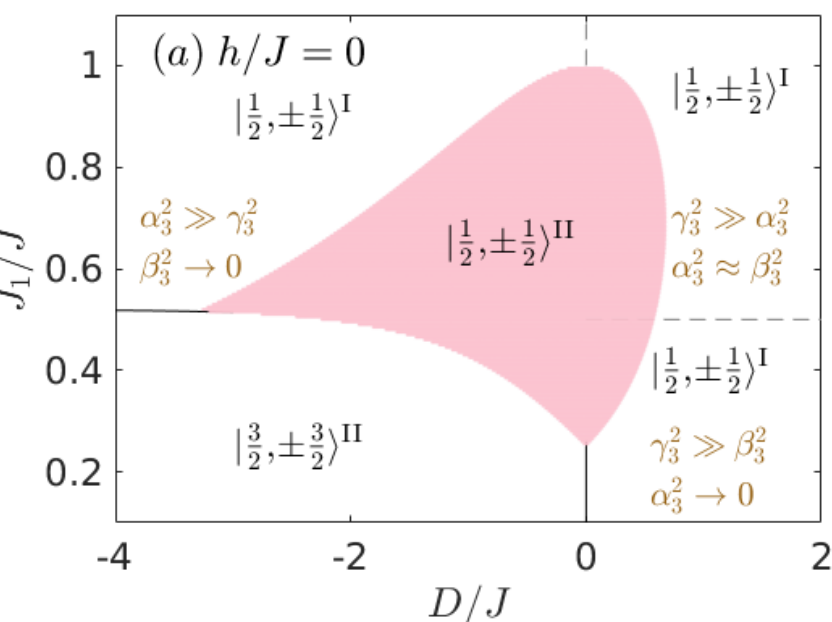}}
{\includegraphics[width=0.33\textwidth,height=4.39cm,trim=3.cm 8.5cm 4.4cm 8.9cm, clip]{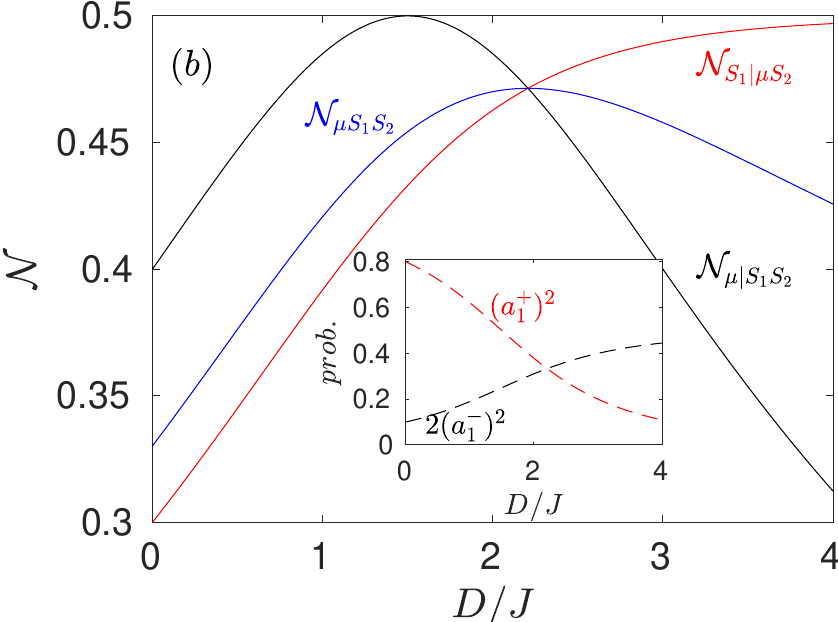}}
{\includegraphics[width=0.33\textwidth,height=4.39cm,trim=3.cm 8.5cm 4.4cm 8.9cm, clip]{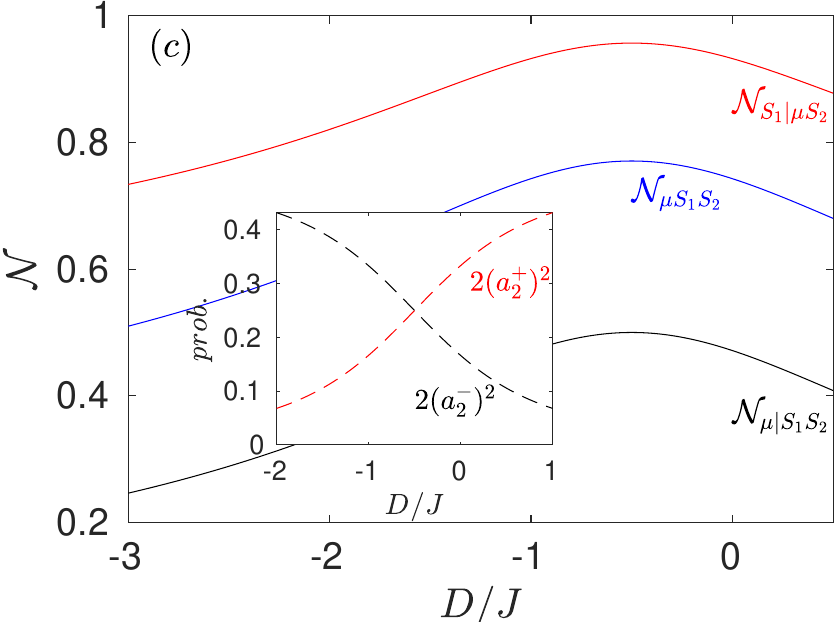}}
\caption{($a$) Stability regions of ground states at $h/J\!=\!0$. The gTN of the phase ($b$) $\vert \tfrac{3}{2},\tfrac{3}{2}\rangle^{\rm II}$ and ($c$) $\vert \tfrac{1}{2},\tfrac{1}{2}\rangle^{\rm II}$. Insets: Behavior of coefficients $(a_1^{\pm})^2$~and~$(a_2^{\pm})^2$.}
\label{fig2}
\end{figure*}

Fig.~\ref{fig2}($a$) illustrates the regions of existence for the phase $\vert \tfrac{1}{2},\pm\tfrac{1}{2}\rangle^{\rm I}$ at $h/J\!=\!0$, where each ground state is two-fold degenerate with respect to the sign of $S_t^z$. Beyond $D/J\!=\!0$, the $\vert \tfrac{1}{2},\pm\tfrac{1}{2}\rangle^{\rm I}$ state occupies a wide parameter space, but the character of the eigenvector differs in each quadrant, separated by the gray dashed lines. Increasing the magnitude of $D/J\!<\!0$ progressively suppresses the maximally entangled spin-$1$ dimer, favoring collinear alignment of all three spins. For $D/J\!>\!0$, the system favors spin-$1$ projections in the $x$-$y$ plane when $J_1/J\!=\!0.5$, whereas for $J_1/J\!>\!0.5$,  collinear basis states become increasingly dominant at the expense of those with non-collinear orientations within the spin-$1$ dimer.

Now, let us examine the gTE in more detail. 
The degree of gTE in the $\vert \tfrac{3}{2},\tfrac{3}{2}\rangle^{\rm II}$ phase - interpreted as a superposition of a collinear basis state $\vert \!-\!1/2\rangle\otimes\vert1,1\rangle$ with a fully separable spin-$1$ dimer, and a symmetric state $\vert 1/2\rangle\otimes\vert\Phi_S^+\rangle$ with an entangled spin-$1$ dimer - is independent of both $J_1$ and $h$, reaching a maximum value  of $\sqrt{2}/3$ at $D/J\!\approx\!2.21$. Below this point, the collinear separable component dominates over the symmetric contribution to the  $\vert \tfrac{3}{2},\tfrac{3}{2}\rangle^{\rm II}$ state, leading to an enhancement of ${\cal N}_{\mu|S_1S_2}$. The maximum value  ${\cal N}^{max}_{\mu|S_1S_2}\!=\!1/2$ is reached at a slightly lower anisotropy of $D/J\!=\!3/2$ (Fig.~\ref{fig2}($b$)).
For $D/J\!\gtrsim\! 2.21$, the anisotropy term $D/J$ increasingly favors the symmetric component of the $\vert \tfrac{3}{2},\tfrac{3}{2}\rangle^{\rm II}$  state, characterized by partial entanglement of the spin-$1$ dimer. While  ${\cal N}_{S_1|\mu S_2}$ gradually saturates at $1/2$ with increasing $D/J$, the gTE diminishes due to a significant reduction in ${\cal N}_{\mu|S_1S_2}$. 

Similarly, the gTN of the  $\vert \tfrac{1}{2},\tfrac{1}{2}\rangle^{\rm II}$ phase - interpreted as a superposition of two antisymmetric states, $\vert 1/2\rangle\otimes\vert\Psi^-\rangle$ and $\vert\!\!-\!\!1/2\rangle\otimes\vert\Phi_A^-\rangle$ - is once again governed solely by the $D/J$ ratio. The maximum gTN of 0.771 is achieved at $D/J\!=\!-1/2$ (Fig.~\ref{fig2}($c$)). Below this point, the collinear component of the spin-$1$ dimer dominates the global state, while the non-collinear component becomes more significant as $D/J$ increases further. 
In contrast to the behavior observed in the $\vert \tfrac{3}{2},\tfrac{3}{2}\rangle^{\rm II}$ phase, both bipartite negativities, ${\cal N}_{\mu|S_1S_2}$ and ${\cal N}_{S_1|\mu S_2}$, exhibit symmetric dependence on $D/J$, with their  maxima located at the same value, $D/J\!=\!\!-1/2$. However,  ${\cal N}_{S_1|\mu S_2}$ consistently exceeds ${\cal N}_{\mu|S_1S_2}$ in magnitude across the entire parameter range. 

One of the most interesting observations of our study is illustrated in Fig.~\ref{fig1}($c$). As shown in our previous work~\cite{Vargova2024}, a strong exchange coupling between identical spin-$1$ particles ($J_1/J\!>\!1$) leads to a biseparable ground state of the form $\vert 1/2\rangle\otimes\tfrac{1}{\sqrt{3}}(\vert 1,\!-\!1\rangle\!+\!\vert \!-\!1,1 \rangle\!-\!\vert 0,0\rangle)$, with maximal entanglement within the spin-$1$ dimer. Naturally, in the  $D/J\!=\!0$ case, the gTN defined as the geometric mean of all bipartite contributions - vanishes. As established,  the non-zero anisotropy term  $D/J\!\neq\!0$  alters the weight of the basis states $\vert1/2\rangle\otimes\vert\pm1,\mp1\rangle$ and $\vert1/2\rangle\otimes\vert0,0\rangle$ (deviating from the uniform $1/\sqrt{3}$), thereby inducing entanglement within the spin-($1/2$,$1$) dimer. Consequently, the ground state $\vert \tfrac{1}{2},\tfrac{1}{2}\rangle^{\rm I}$ is slightly modified, as shown in Tab.~\ref{tab1}.  A direct consequence of these changes is the onset of  gTE for  arbitrarily small $D/J\!\neq\!0$. It is important to emphasize, that unlike the previously discussed ground states the probability coefficients $\alpha_3^2$, $\beta_3^2$, and $\gamma_3^2$ depend on both $D/J$ and $J_1/J$, with their behavior differing for positive and negative values of $D/J$. For example, at $J_1/J\!=\!1.5$ the gTN reaches a maximum of $0.199$ at $D/J\!=\!1.807$. More generally, the gTN of the $\vert \tfrac{1}{2},\tfrac{1}{2}\rangle^{\rm I}$ state attains its overall maximum value of $0.436$ in the limit $J_1/J\!\rightarrow\!0^+$. This finding is particularly significant, as experimentally realized spin-($1$,$1/2$,$1$) magnetic complexes typically exhibit nearly linear spin structures with a negligible value of the coupling constant $J_1$~\cite{Ribas,Podlesnyak}.

For $J_1/J\!>\!1$, the phase diagram includes an additional ground state, $\vert \tfrac{3}{2},\tfrac{3}{2}\rangle^{\rm I}$. Since this state  is biseparable with respect to the tensor product of the spin-$1/2$ and spin-$1$ dimer, the gTN vanishes in this region of the parameter space. For clarity, the boundary between the $\vert \tfrac{3}{2},\tfrac{3}{2}\rangle^{\rm I}$ and the fully separable $\vert \tfrac{5}{2},\tfrac{5}{2}\rangle$ ground states is indicated by a dashed line in Fig.~\ref{fig1}($c$).

\begin{figure}[h!]
{\includegraphics[width=0.5\columnwidth,trim=3.5cm 8.9cm 5.7cm 8.7cm, clip]{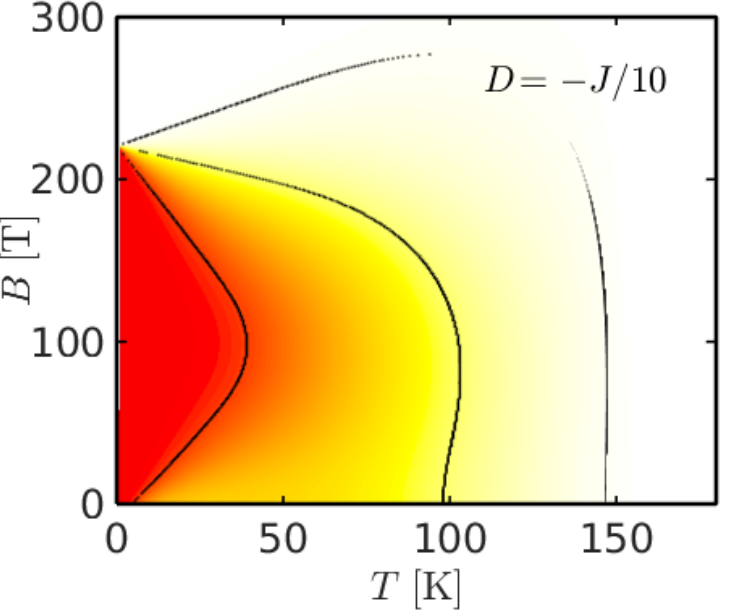}}{\includegraphics[width=0.515\columnwidth,trim=5.3cm 8.9cm 3.6cm 8.7cm, clip]{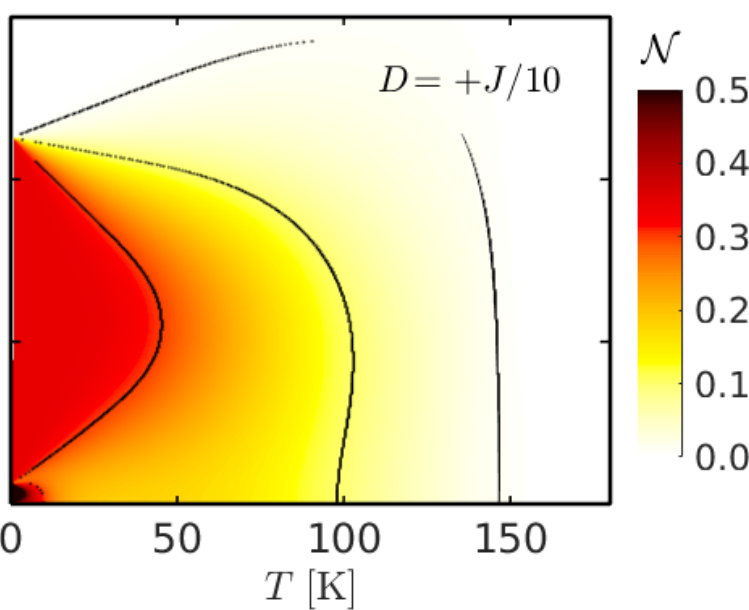}}
\caption{Density plot of ${\cal N}_{\mu S_1S_2}$ for the Ni$_2$Cu compound with parameters $J/k_B\!=\!90.3$ cm$^{-1}$, $J_1/k_B\!=\!0$ cm$^{-1}$, $D/J\!=\!\pm 10$, and $g\!=\!(2g_{\rm Ni}\!+\!g_{\rm Cu})/3$, adapted from Ref.~\cite{Ribas}. Black contour lines indicate isovalues of $0.3$, $0.1$, and $0.01$ (from left to right).
}
\label{fig3}
\end{figure}

An important aspect of our study is to examine whether $D/J$ significantly affects the thermal stability of gTE. Fig.~\ref{fig3} presents theoretical predictions for thermal gTE in a real heterobimetallic complex NiCuNi~\cite{Ribas}.
Since the value of $D/k_B$ could not be determined from previous magnetic measurements due to weak sample sensitivity~\cite{Ribas}, we performed simulations using both positive and negative values of $D/k_B$, each with a magnitude equal to one-tenth of $J/k_B$. Comparing these results to the $D/k_B\!=\!0$ case~\cite{Vargova2024}, we surprisingly find that $D/k_B$ has a negligible impact on the thermal stability of gTE. However, both zero-temperature entanglement value and its robustness under applied magnetic fields show slight enhancement with increasing $D/k_B$. Thus, we conclude that $D/k_B$ is not a decisive factor in determining the thermal stability of gTE in the mixed spin-($1$,$1/2$,$1$) Heisenberg trimer.

Last but not least, let us turn our attention to the case $J_1/J=1.5$. As previously identified for the $D/J=0$ limit,  thermal fluctuations play a key role in stimulating gTE with relatively strong magnitude~\cite{Vargova2024}. This behavior is primarily driven by the thermal population of the low-lying excited state $\vert \tfrac{1}{2},\tfrac{1}{2}\rangle^{\rm II}$, which is entangled with respect to the spin-$1/2$ and spin-$1$ dimer and contributes to the density matrix elements $\rho_{3,11}^-$ and $\rho_{3,13}^-$, as shown in Eqs.~(D.11) of Ref.~\cite{Vargova2024}. As a result, the negative eigenvalue $(\lambda_5^-)_{\mu}$  of the matrix $\mathbf{Q}_3^{\mu}(-)$ (Eqs.~(E.12), (E.13) in Ref.~\cite{Vargova2024}) gives rise to  a non-zero value of ${\cal N}_{\mu|S_1S_2}$.
An arbitrary single-ion anisotropy $D/J\!\neq\!0$ destroys the biseparability of the  $\vert \tfrac{1}{2},\tfrac{1}{2}\rangle^{\rm I}$ ground state by inducing a superposition with additional states including non-collinear orientations of the identical spin-$1$ particles in the corresponding eigenvector.
\begin{figure}[h!]
\;{\includegraphics[width=7.1cm,height=4.6cm,trim=3.cm 9.9cm 4.cm 8.5cm, clip]{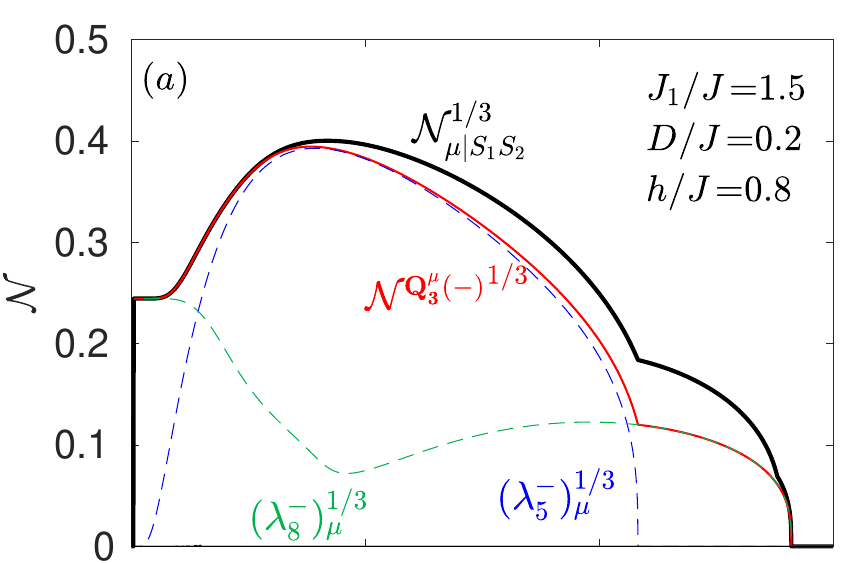}}
\\
{\includegraphics[width=8.5cm,height=5.1cm,trim=3.4cm 9cm 3.2cm 8.5cm, clip]{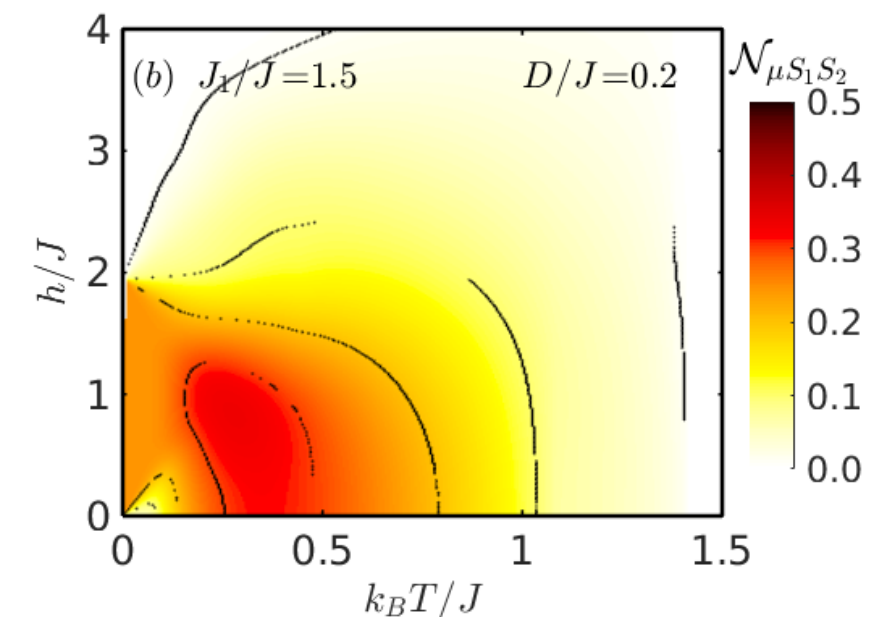}}
\caption{($a$) Thermal contributions of $(\lambda_5^-)_{\mu}^{1/3}$
and $(\lambda_8^-)_{\mu}^{1/3}$ to the bipartite negativity ${\cal N}_{\mu|S_1S_2}^{1/3}$, along with the contribution of ${({\cal N}^{\mathbf{Q}_3^{\mu}(-)})}^{1/3}$~\cite{Vargova2024}. ($b$)  Thermal behavior of gTN as a function of $h/J$ at $J_1/J\!=\!1.5$ and $D/J\!=\!0.2$, with isolines at $0.3$, $0.2$, $0.1$ and $0.01$.
}
\label{fig4}
\end{figure}
As a consequence, the matrix elements  $\rho_{3,11}^-$, $\rho_{3,13}^-$, and $\rho_{5,11}^-$, which determine the eigenvalues of $\mathbf{Q}_3^{\mu}(-)$, become non-zero at $k_BT/J\!=\!0$. Nevertheless, the thermal behavior of $(\lambda_5^-)_{\mu}$ remains nearly unchanged due to the consistent contribution of the  $\vert \tfrac{1}{2},\tfrac{1}{2}\rangle^{\rm I}$ ground state (Fig.7($a$) in Ref.~\cite{Vargova2024}). On the other hand, the element $\rho_{5,11}^-$ immediately activates the eigenvalue $(\lambda_8^-)_{\mu}$  (Eq.~(E.13) in Ref.~\cite{Vargova2024}), which fully determines the low-temperature behavior of the gTN (Fig.~\ref{fig4}($a$)).
Similar to the $D/J\!=\!0$ case, the gTN at $J_1/J\!=\!1.5$ exhibits a notable singularity above $k_BT/J\!=\!1$, stemming from  the same mechanism, namely, a sign change  of an eigenvalue of the partially transposed density matrix from negative to positive. The thermal dependence of ${\cal N}_{\mu S_1S_2}$ for various values of $h/J$ and a representative value $D/J\!=\!0.2$ is illustrated in Fig.~\ref{fig4}($b$).

\section{Conclusions}\label{AA}
To conclude, our study reveals that single-ion anisotropy  plays a crucial role in shaping the gTE in the mixed spin-($1$,$1/2$,$1$) Heisenberg trimer by significantly influencing the structure of the eigenvectors. The variation of weight coefficients with $D/J$ induces a non-uniform dependence of the gTN, with the sole exception of the $\vert \tfrac{3}{2},\tfrac{3}{2}\rangle^{\rm I}$ phase, in which the gTE remains unaffected.
In the physically relevant limit of $J_1/J\!\to\!0$, the gTE increases steadily with increasing anisotropy. Conversely, for $J_1/J\!=\!1.5$, even a small non-zero anisotropy term $D/J$ immediately disrupts the biseparable ground state and induces entanglement for both signs of $D/J$. Furthermore, a finite value of $D/J$ favors the stabilization of the $\vert \tfrac{1}{2},\tfrac{1}{2}\rangle^{\rm I}$ ground state across a wide range of $J_1/J$, in contrast to the $D/J\!=\!0$ case, where this state appears only at $J_1/J\!\geq\!1$.
Despite its pronounced effect on the ground-state properties, $D/J$ exerts only a minor influence on the thermal robustness of the globally entangled state, as demonstrated for parameter sets relevant to the real NiCuNi complex~\cite{Ribas}. Finally, we have shown that the non-trivial temperature dependence of the global negativity at $J_1/J\!=\!1.5$ arises from the thermal activation of additional non-collinear basis states in the ground-state~eigenvector.
\section*{Acknowledgment}
This work was financially supported under the grant Nos.VEGA 1/0298/25 and APVV-20-0150.



\end{document}